\documentclass[conference]{IEEEtran}
\IEEEoverridecommandlockouts
\usepackage{cite}
\usepackage{amsmath,amssymb,amsfonts}
\newtheorem{myrem}{Remark}

\usepackage{algorithmic}
\usepackage{graphicx}
\usepackage{textcomp}
\usepackage{xcolor}
\def\BibTeX{{\rm B\kern-.05em{\sc i\kern-.025em b}\kern-.08em
    T\kern-.1667em\lower.7ex\hbox{E}\kern-.125emX}}
\begin{document}

\title{A Pricing Rule for Third-Party Platoon Coordination Service Provider}

\author{Ting~Bai,
        Alexander~Johansson,
        Shaoyuan~Li
        and~Jonas~Mårtensson% <-this % stops a space
\thanks{Ting Bai, Alexander Johansson and Jonas Mårtensson are with the Integrated Transport Research Lab, Digital Futures, and Division of Decision and Control Systems, School of Electrical Engineering and Computer Science, KTH Royal Institute of Technology, 100 44 Stockholm, Sweden (e-mail: tingbai@kth.se; alexjoha@kth.se; jonas1@kth.se).}% <-this % stops a space
\vspace{2pt}

\thanks{Shaoyuan Li is with the Key Laboratory of System Control and Information Processing, Ministry of Education, and the department or Automation, Shanghai Jiao Tong University, 200240, Shanghai, China (e-mail: syli@sjtu.edu.cn)}
\thanks{Manuscript received January 30, 2022.}}

%\author{\IEEEauthorblockN{1\textsuperscript{st} Ting Bai}
%\IEEEauthorblockA{\textit{Division of Decision and Control Systems} \\
%\textit{KTH Royal Institute of Technology}\\
%Stockholm, Sweden\\
%tingbai@kth.se}
%\and
%\IEEEauthorblockN{2\textsuperscript{nd} Shaoyuan Li}
%\IEEEauthorblockA{\textit{Department of Automation} \\
%\textit{Shanghai Jiao Tong University}\\
%Shanghai, China\\
%syli@sjtu.edu.cn}
%\and
%\IEEEauthorblockN{3\textsuperscript{th} Karl Henrik Johansson}
%\IEEEauthorblockA{\textit{Division of Decision and Control Systems} \\
%\textit{KTH Royal Institute of Technology}\\
%Stockholm, Sweden\\
%kallej@kth.se}
%\and
%%\IEEEauthorblockN{4\textsuperscript{th} Jonas Mårtensson}
%\IEEEauthorblockA{\textit{Division of Decision and Control Systems} \\
%\textit{KTH Royal Institute of Technology}\\
%Stockholm, Sweden\\
%jonas1@kth.se}
%}

\maketitle

\begin{abstract}
We model a platooning system including trucks and a third-party service provider that performs platoon coordination, distributes the platooning profit within platoons, and charges the trucks in exchange for its services. This paper studies one class of pricing rules, where the third-party service provider keeps part of the platooning profit each time a platoon is formed. Furthermore,  we propose a platoon coordination solution based on distributed model predictive control in which the pricing rule is integrated. To evaluate the effect of the pricing on the platooning system, we perform a simulation over the Swedish road network. The simulation shows that the platooning rate and profit highly depend on the pricing. This suggests that pricing needs to be set carefully to obtain a satisfactory platooning system in the future.
\end{abstract}
\vspace{4pt}

\vspace{5pt}

\section{Introduction}
Platooning is when trucks are lined up in a formation on the road with small distances between the trucks. Platooning technology is enabled by sensor, communication, and automation technologies, that have been taking great leaps over the last decades. In comparison to normal driving, platooning improves the aerodynamics which leads to reduced fuel consumption, especially for follower trucks in a platoon. For example, experimental results in~\cite{bonnet2000fuel} show that follower trucks can achieve fuel savings of up to $21\%$ with $10$m inter-vehicular distances and a speed of $80$~km/h. The experimental results in~\cite{janssen2015truck} show that truck platooning yields fuel consumption reduction of $8$-$13\%$ for the follower truck in a platoon comprised of two trucks. In addition to energy savings, platooning has the potential to increase the road capacity, reduce green house gas emission and the workload of drivers, as reported in~\cite{tsugawa2016review,massar2021impacts,conlon2019greenhouse,maiti2019impact}. 

Due to the aforementioned benefits, platooning has been researched on by trucks manufactures, freight companies and academia. The first research project on truck platooning started in the 1990s in the European Commission’s CHAUFFEUR project~\cite{benz1996telematics}, and has been followed by a series of research projects from all around the world, for example, the California PATH program~\cite{shladover1992california}, the Energy ITS project~\cite{tsugawa2013overview} in Japan, the European SARTRE~\cite{robinson2010operating} project, and so on~\cite{van2012cooperative}. The main focus of the platooning research has been on safe and efficient control of platoons, see e.g.,~\cite{besselink2017string,jia2016platoon,xiao2011practical}. 

Trucks with different routes and time schedules need to be coordinated to fully reap the platooning benefits. By coordination, trucks can expect to join more platoons for fuel economy through the way of, for example, planning their routes~\cite{luo2018coordinated}, adjusting driving speeds~\cite{van2017efficient}, changing waiting or departure times at hubs~\cite{zhang2017freight}. Most previous work on platoon coordination study centralized schemes where the aim is to optimize the overall platooning profit of all trucks. However, this is not applicable when trucks are owned by different carriers since they in practice are interested to optimize their individual profits.

The authors in~\cite{johansson2021strategic} developed a non-cooperative game-theoretic framework to study the strategic interaction among trucks when they decide on their waiting times at hubs to form platoons. The efficiency of this algorithm is limited because a large number of iterations may be required to find a Nash equilibrium in the platoon coordination game when the number of trucks is large. Another work considering platoon coordination of trucks with individual utility functions is~\cite{bai2021event}. In this work, a distributed model predictive control method is presented as solution wherein trucks plan their waiting times independently while relying on the predicted arrival times of others. The aforementioned works assume that trucks are willing to communicate their routes and schedules with other trucks in the network. In practice, trucks from different carriers may not communicate directly due to privacy and trust concerns. Then, one solution to avoid direct communication among trucks is to employ a third-party service provider that communicates with trucks and performs coordination. 

In a platooning system with a third-party service provider, the commercial attribute of the service provider makes the profit sharing between trucks and the provider a new problem. The works in \cite{8917349} and~\cite{sun2019behaviorally} studied profit-sharing schemes within platoons, where \cite{8917349} proposes three different profit-sharing schemes and models trucks' behavior in a non-cooperative game. On the contrary, the authors in~\cite{sun2019behaviorally} model the trucks' behavior in a cooperative game, where the presented profit-sharing scheme makes trucks do not have any incentive not to follow the system-optimal solution. One difference between these works and our work is that a large-scale system is considered in our work while~\cite{8917349} and~\cite{sun2019behaviorally} consider small-scale systems. Another difference is that the profit-sharing between trucks and the service provider is taken into account in our work. 

The objective of this paper is to propose and evaluate a pricing rule for the third-party service provider which offers real-time communication and platoon coordination services to trucks, and charges amount of service fee in return. In this paper, we present a pricing rule in which the service provider keeps a proportion of the platooning profit and sets the service fee to charge trucks, making the profit is evenly distributed among trucks. In addition, we integrate the pricing rule with our platoon coordination solution developed in~\cite{bai2021event}, which enables the service provider to provide an optimal suggested waiting time to each truck with considering its platooning reward, waiting cost and service fee. The simulation evaluates the pricing rule and shows its effect on the profit distribution between the service provider and trucks. The main contributions of this paper are threefold: 
\vspace{1pt}

\begin{itemize}
    \item [(i)] We model a transportation system with a third-party service provider and propose a pricing rule for charging trucks that use the platoon coordination service;
    \vspace{2pt}
    
    \item [(ii)] We propose a platoon coordination method, based on the distributed MPC, in which the pricing rule is integrated;
    \vspace{2pt}
    
    \item [(iii)] The pricing rule is evaluated in the simulation over the Swedish road network with hundreds of trucks. 
\end{itemize}
\vspace{2pt}

The remainder of the paper is organized as follows. Section~II formally presents the pricing rule for the service provider, including the charging and compensation rules. In Section~III, we introduce how to integrate the pricing rule with the distributed MPC platoon coordination method. Section~IV provides the simulation and evaluation results. Finally, Section~V concludes this paper with the a brief conclusion and points out the directions for the future work. 
\vspace{3pt}

\section{Pricing Rule of the Third-Party Service Provider}
\begin{figure}[thpb]
     \centering
     \includegraphics[width=0.95\linewidth]{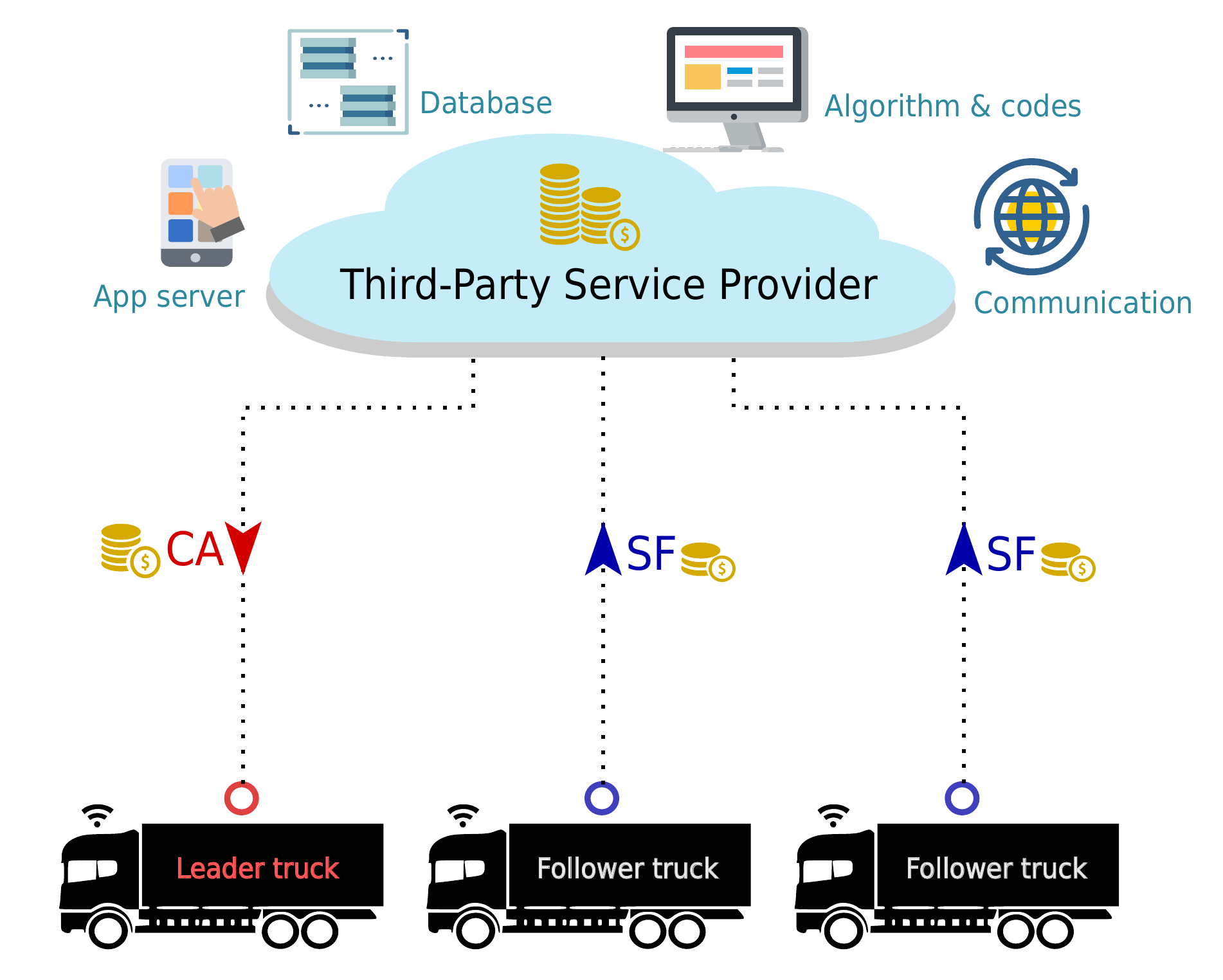}
      \caption{The financial flow between the third-party service provider and trucks, where CA and SF stand for the compensation allowance and service fee, respectively.}
      \vspace{-2pt}
      \label{Fig.1}
   \end{figure}

This section presents the pricing rule for a platooning system consisting of trucks and a third-party service provider. Fig. \ref{Fig.1} shows the financial flow between the third-party service provider and trucks. The service provider collects trucks' information and performs platoon coordination. Moreover, it distributes the platooning profit within platoons and keeps some profit for offering the services. 
As previously mentioned, the fuel savings due to platooning are different for leader trucks and its follower trucks. To distribute the platooning benefit evenly among trucks, we intend to charge the follower trucks a service fee and compensate the leader truck each time a platoon is formed. First, the charging rule for the follower trucks is proposed. Then, the compensation rule for the leader trucks is proposed.   

\subsection{Charging rule}
 Consider a platoon of $n$ trucks where each truck focuses on its own platooning benefit. According to experimental results on truck platoons, see e.g., \cite{bishop2017evaluation}, \cite{mcauliffe2017fuel}, each follower truck in a platoon has approximately the same fuel consumption reduction while the leader truck has significantly lower fuel saving. Therefore, we assume that every follower truck enjoys the same platooning benefit and the leader truck has no benefit. In this work, we propose a charging rule where a service fee is charged every follower truck and the leader truck instead receives a compensation. The service fee and compensation is set to even out the imbalance in profit and the service provider keeps part of the profit.
\vspace{2pt}

We denote by $P_f$ the platooning benefit of a follower truck obtained by forming platoons with other trucks. The service fee to be paid to the service provider is represented by $F_f$, which includes two parts: one is the profit $R_{s,f}$ kept by the service provider and the other part is the compensation $R_{c,f}$ used to charge the leader truck. To guarantee that the leader and follower trucks share the rest platooning profit, i.e., $(n\!-\!1)(P_f\!-\!R_{s,f})$, equally, each truck in the platoon should keep the profit of 
\begin{align}
    \bar{R}_t=\frac{(n\!-\!1)(P_f\!-\!R_{s,f})}{n}.\label{Equ.1}
\end{align}
Since the profit to compensate the leader truck is shared among all follower trucks, each follower truck spends $R_{c,f}\!=\!\bar{R}_t/(n\!-\!1)$ on the compensation. Thus, the service fee of every truck is denoted by 
\begin{align}
    F_f&=R_{s,f}+R_{c,f}\label{Equ.2}\\
    &=R_{s,f}+\frac{(P_f\!-\!R_{s,f})}{n},\label{Equ.3}
\end{align}
where $R_{c,f}$ is the compensation contributed by each follower truck in a platoon to charge the leader truck. Sum the service fees taken from every follower truck in the platoon, the total profit kept by the service provider from one platoon is 
\begin{align}
    F=(n\!-\!1)R_{s,f}.\label{Equ.4}
\end{align}

\subsection{Compensation rule}
The leader truck in a platoon receives a compensation to even out the benefit between the leader and its follower trucks. That is, given a platoon with size $n$, the total compensation that the leader truck can receive, denoted by $R_c$, is of the form
\begin{align}
    R_c=(n\!-\!1)R_{c,f}=\bar{R}_t\label{Equ.5}
\end{align}
which equals to the platooning profit $R_f$ saved by each follower truck in the platoon after paying for its service fee, i.e., 
\begin{align}
    R_f&=P_f-F_f=R_c.\label{Equ.6}
\end{align}
In this way, every truck in the platoon achieves the same platooning benefit, which is related with the platooning reward and service fee of each follower, as well as the number of trucks in the platoon. 
\vspace{3pt}

\begin{myrem}
According to Equ.~(\ref{Equ.4}), the service provider is able to regulate its total profit by changing the value of $R_{s,f}$, which in turn affects every truck's profit by Equ.~(\ref{Equ.1}) and every truck's willingness to use the platoon coordination service. In this sense, there should be a trade-off between the profits of the service provider and trucks in the platooning system. 
\end{myrem}
\vspace{5pt}

\section{Platoon Coordination}

This section introduces how the service provider offers the platoon coordination service to individual trucks with taking into account the charging and compensation rules. First, we give the system model which includes the road network and the dynamical models of trucks. Then, the utility of individual trucks is proposed. Subsequently, we provide the optimization problem that is used to compute the optimal platoon coordination solutions to trucks. 
\vspace{2pt}

\subsection{System model}
We consider a large-scale transportation system that consists of $M$ trucks, where every truck travels from its origins to the destinations to fulfill a delivery task. Each truck $i\!\in\!{\mathcal{M}\!=\!\{1,2,\dots,M\}}$ has a fixed route and can form platoons at the hubs alongside its route. By indexing the origin as the first hub and the destination as the $N_i$-th hub, the route of truck $i$ can be represented by the set
\begin{align}
    \mathcal{E}_i=\{e_i(1),e_i(2),\dots,e_i(N_i\!-\!1)\},\label{Equ.7}
\end{align}
where $e_i(k)$ denotes the road segment connecting the $k$-th and $(k\!+\!1)$-th hub of truck $i$.

We use $a_i(k)$ and $w_i(k)$ to denote the arrival time and waiting time of truck $i$ at its $k$-th hub, respectively. The travel time on the route segment $e_i(k)$ is denoted by $c_i(k)$. Then, the arrival times at hubs of truck $i$ can be computed according to the following equations
\begin{align}
    &a_i(1)=t_{i,start}\label{Equ.8}\\
    &a_i(k\!+\!1)=a_i(k)\!+\!w_i(k)\!+\!c_i(k),~k\!=\!1,2,\dots,N_i\!-\!1,\label{Equ.9}
\end{align}
where truck $i$'s arrival time at its first hub (i.e., the origin) in its route is represented as $t_{i,start}$. Seeing that every trip has its delivery deadline which needs to be respected, we require that $a_i(N)\!\leq\!{t_{i,end}}$, where $a_i(N_i)$ is the arrival time of truck $i$ at its destination and $t_{i,end}$ is the deadline.  
\vspace{2pt}

\subsection{Utility of trucks}
The utility of every truck includes the reward it gains from forming platoons and its loss caused by waiting at hubs. 
\vspace{3pt}
\subsubsection{Reward function} Let us recall Equ. (\ref{Equ.5}) and (\ref{Equ.6}). For any truck $i$ traveling in a platoon on its $k$-th route segment $e_i(k)$, no matter it is a follower or a leader truck, its platooning profit is the same, and $P_f$ can be described as the form
\begin{align}
    P_f=\xi_ic_{i}(k),\label{Equ.10}
\end{align}
where $\xi_i$ denotes the platooning profit per follower truck and per travel time unit. Additionally, $R_{s,f}$ is the portion of the service fee retained by the service provider, which is considered as the following form  
\begin{align}
    R_{s,f}=\alpha{P_f},\label{Equ.11}
\end{align}
where $0\!\leq\!\alpha\!\leq\!{1}$ is the adjustable parameter to regulate the profit sharing between the service provider and trucks. 
\vspace{2pt}

To form a platoon, the departure times and the routes of the trucks in the platoon must be synchronized. Subsequently, the size of a platoon is characterized. The dynamic model of individual trucks in Equ.~(\ref{Equ.9}) indicates that the arrival time of a truck at its next hub is predictable given its arrival and waiting times at the current hub. In other words, with the knowledge of other trucks' predicted arrival times and waiting times at hubs, the service provider is able to predict the platooning partners for every truck at each of its hubs. 
\vspace{2pt}

Let $\mathcal{R}_i(k\!+\!h|k)$ denote the predicted platooning partners of truck $i$ on its $(k\!+\!h)$-th route segment, which is predicted by the service provider to decide truck $i$'s waiting times at its $k$-th hub. More precisely, $\mathcal{R}_i(k\!+\!h|k)$ consists the trucks that are predicted to depart from the $(k\!+\!h)$-th hub on $\mathcal{E}_i$ at the same time as truck $i$ and have the next route segment in common, including truck $i$ itself. Mathematically, a truck $j\!\in\!{\mathcal{M}}$ that has $e_i(k\!+\!h)$ in its route is part of the set $\mathcal{R}_i(k\!+\!h|k)$ if
\begin{align}
   a_i(k\!+\!h|k)+w_i(k\!+\!h|k)=d_{j,i}(k\!+\!h|k),\label{Equ.12}
\end{align}
where $d_{j,i}(k\!+\!h|k)$ denotes the predicted departure time of truck $j$ at the $(k\!+\!h)$-th hub of truck $i$ when it makes the decision at its $k$-th hub. By this definition, the size of truck~$i$'s platoon formed on the route segment $e_i(k\!+\!h|k)$ equals to the cardinality of its platooning partners, i.e., $n\!=\!|\mathcal{R}_i(k\!+\!h|k)|$. 
\vspace{2pt}

In consequence, by Equ.~(\ref{Equ.5}), (\ref{Equ.10}) and (\ref{Equ.11}), the platooning benefit of truck $i$ on its route segment $e_i(k\!+\!h)$ predicted at its $k$-th hub is denoted by
\begin{align}
    R_i(k\!+\!h|k)=(1\!-\!\alpha)\xi_ic_i(k)\frac{|\mathcal{R}_i(k\!+\!h|k)|\!-\!1}{|\mathcal{R}_i(k\!+\!h|k)|}\label{Equ.13}
\end{align}
with $\mathcal{R}_i(k\!+\!h|k)$ denoting the set of trucks in the platoon.
\vspace{1pt}

Considering all the platooning benefits of truck $i$ on its remaining route segments, i.e., from its $k$-th to $(N_i\!-\!1)$-th route segment, the profit that truck $i$ predicts to achieve is 
\begin{align}
    R_i(k)=\sum_{h=0}^{N_i-1-k}R_i(k\!+\!h|k).\label{Equ.14}
\end{align}
This reward function of truck $i$ is used by the service provider to coordinate platoons in the platooning system and schedule the waiting times of truck $i$ at its hubs.
\vspace{4pt}

\subsubsection{Loss function}
Waiting at hubs could increase the probability of forming platoons with others for individual trucks, but may also increase their waiting loss due to a delayed delivery of goods or higher costs for drivers. For any truck $i$ at its $k$-th hub, the predicted waiting cost at all the remaining hubs between its $k$-th and $(N_i\!-\!1)$-th hub is denoted as
\begin{align}
    L_i(k)=\sum_{h=0}^{N_i-1-k}\epsilon_iw_i(k\!+\!h|k),\label{Equ.15}
\end{align}
where $\epsilon_i$ represents the monetary loss per time unit and $w_i(k\!+\!h|k)$ denotes the waiting time of truck $i$ at its $(k\!+\!h)$-th hub predicted at the $k$-th hub.
\vspace{2pt}

Taken the above reward and loss functions together, the utility of truck $i$ at its $k$-th hub is denoted as
\begin{align}
    J_i(k)=R_i(k)\!-\!L_i(k).\label{Equ.16}
\end{align}
By maximizing such a utility for ever truck arriving at a hub, the service provider is able to provide an optimal suggested waiting time for trucks to form platoons.

\vspace{1pt}

\subsection{Platoon coordination problem}
Given the route information, dynamical models and time constraints of every truck, the service provider is able to provide an optimal suggested waiting time at a hub for every truck. Based on the distributed MPC method, the optimal platoon coordination solution of any truck $i$ at its $k$-th hub can be computed by solving the following problem 
\begin{subequations}
\begin{align}
  \max_{\textit{\textbf{w}}_i(k)}~~ &J_i(k)\label{Equ.17a}\\
  \mathrm{s.t.}~~ &a_i(k|k)=t_{i,arr}(k)\label{Equ.17b}\\
  &a_i(k\!+\!h\!+\!1|k)=a_i(k\!+\!h|k)+w_i(k\!+\!h|k)\label{Equ.17c}\\
  &\quad\quad\quad\quad\quad\quad\quad +c_i(k\!+\!h),~h\!=\!0,1,\dots,N_i\!-\!1\!-\!k\nonumber\\
  &a_i(N_i|k)-t_{i,end}\leq{0},\label{Equ.17d}
  \end{align}
\end{subequations}
where, as previously defined, $J_i(k)$ includes the predicted platooning reward and waiting cost of truck $i$ at its remaining hubs. It is a function of truck $i$'s waiting times and the predicted departure times of other trucks at the hubs $\{k,k\!+\!1,\dots,N_i\!-\!1\}$. The optimization variable $\textbf{\textit{w}}_i(k)$ includes truck $i$'s waiting times at all its remaining hubs, namely,
\begin{align}
    \textbf{\textit{w}}_i(k)=[w_i(k|k),w_i(k\!+\!1|k),\dots,w_i(N_i\!-\!1|k)].\label{Equ.18}
\end{align}
The optimal waiting time $w_i^*(k|k)$ will be provided to truck~$i$ as its optimal suggested waiting time at its $k$-th hub, while the predicted waiting times $[w_i^*(k\!+\!h|k),\dots,w_i^*(N_i\!-\!1|k)]$ and the departure times of truck $i$ will be saved by the third-party service provider when scheduling the waiting times of other trucks. 
\vspace{2pt}

Moreover, the constraint (\ref{Equ.17b}) in the optimization problem sets the arrival time to the current hub $k$ in the predictive model. In Equ.~(\ref{Equ.17c}), $c_i(k\!+\!h)$ is the travel time of truck $i$ on its route segments $e_i(k\!+\!h)$, which is known in advance to the service provider. The last constraint (\ref{Equ.17d}) ensures that truck $i$ respects its delivery deadline at the destination.
\vspace{3pt}

\begin{myrem}
The distributed platoon coordination problem (17) is in nature a mixed integer nonlinear programming (MINP) problem, and we solve it by dynamic programming.  
\end{myrem}
\vspace{3pt}

\begin{myrem}
As the optimal suggested waiting time $w_i^*(k|k)$ is based on the predicted departure times of other trucks in the system, the above platoon coordination method cannot guarantee that truck $i$ will join a platoon by waiting at its $k$-th hub for $w_i^*(k|k)$. Therefore, to establish trucks' acceptability and willingness to use the platoon coordination service, we assume in this paper that the service provider will pay for a truck's waiting loss if it fails to form a platoon with others.      
\end{myrem}
\vspace{5pt}

\section{Simulation and evaluation}
In this section, we evaluate the effect of the pricing rule on the platooning system by conducting simulations over the Swedish road network with hundreds of trucks.   

\subsection{Parameter settings}
In the simulation, we consider a platooning system with hundreds of trucks, where the number of trucks is varied from $100$ to $500$, and the origin and destination pairs (i.e., od pairs) of each truck are randomly selected from $84$ major hubs in the Swedish road network. The route information and travel times on each route segment of trucks are download from \textit{OpenStreetMap}~\cite{OpenStreetMap}. We assume that trucks start their trips at any time between 8:00-12:00 a.m. Each truck travels with the velocity of $80$ kilometers per hour, and the fuel price is $18$~SEK per litre. In line with the guide to EU rules on drivers' hours, the total driving time of each driver per day is less than $9$ hours. Furthermore, the allowed waiting times of every truck at all the hubs in its route are assumed to be within $10\%$ of its total travel times. We assume that the fuel saving of each follower truck in a platoon is $10\%$ of the total fuel consumption. On this basis, the monetary saving of the truck platooning is $0.72$~SEK per follower per kilometer and the parameter $\xi_i$ that captures the platooning benefit is obtained as $57.5$~SEK per follower per hour. According to the labor cost for drivers, every truck's waiting loss at hubs, denoted by $\epsilon_i$, is set as $260$~SEK per hour. 
\vspace{2pt}

The adjustable parameter $\alpha$ in Equ.~(\ref{Equ.11}) is regulated from~$0$~to~$1$ with an interval of $0.1$ to evaluate its influence on the platooning system. For each selection of the parameter $\alpha$, $5$ scenarios are tested, where each scenario includes $100,200,\dots,500$ trucks in the system, respectively. We fix the initial system setting in each scenario, including the od pairs, starting times of different trips, delivery routes, and waiting time constraints, while only change the value of $\alpha$ in the multiple runs.  

\subsection{Evaluation of the pricing rule}
The simulation results are given below. First, Fig.~\ref{Fig.2} shows the total profit of the third-party service provider from the platooning system, which does not include the compensation to charge the leader trucks in platoons. As is shown, the profit of the service provider increases at first then decreases as the parameter $\alpha$ increases. This result indicates that a decreased number of platoons are formed when the service fee is higher than a certain level. In addition, with the same $\alpha$, the more trucks in the system, the more platooning profit the service provider can get.

\begin{figure}[thpb]
     \centering
     \includegraphics[width=0.98\linewidth]{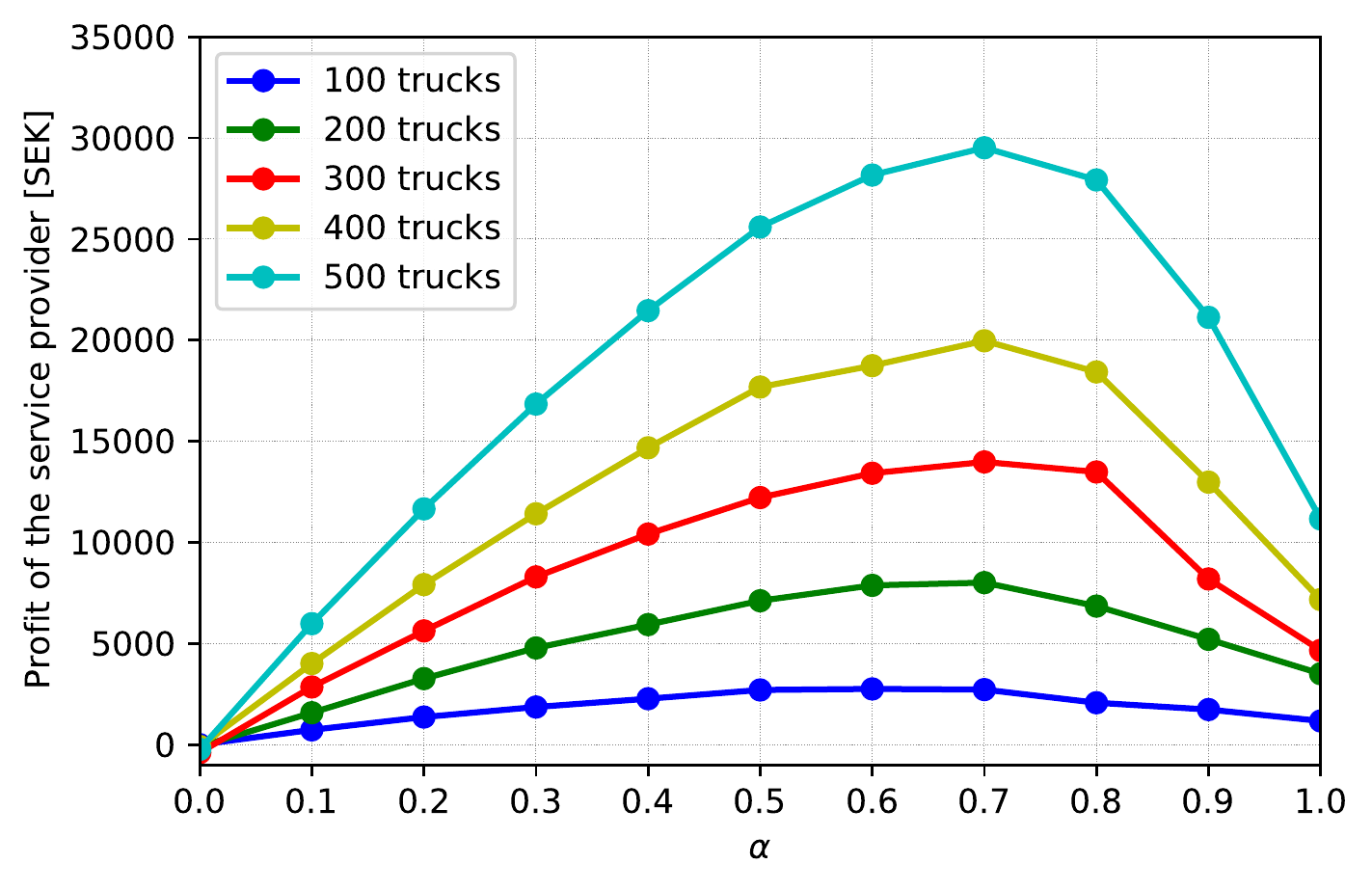}
     \vspace{-3pt}
      \caption{The total profit of the third-party service provider.}
      \label{Fig.2}
   \end{figure}

\begin{figure}[thpb]
     \centering
     \includegraphics[width=0.98\linewidth]{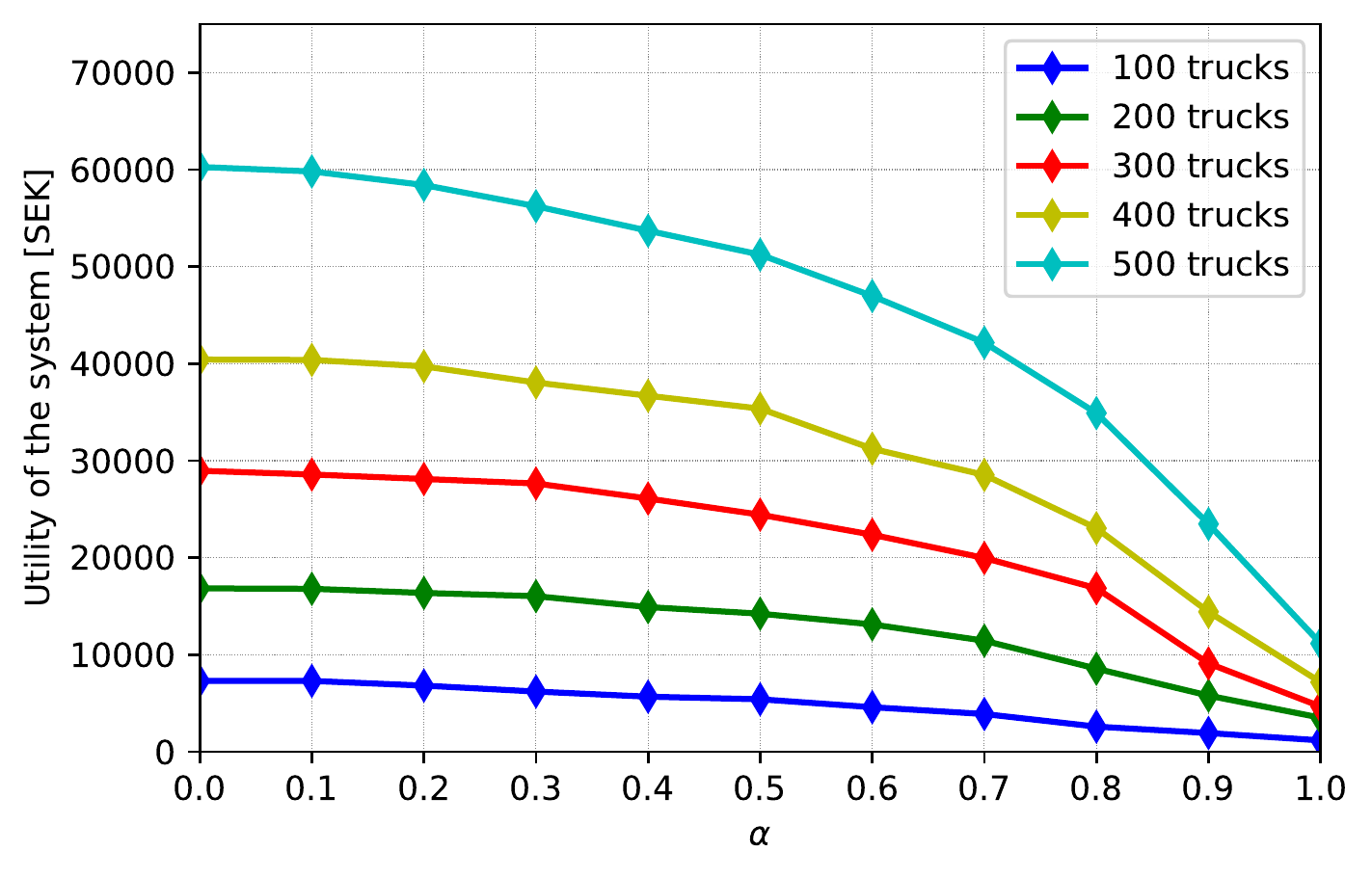}
      \vspace{-3pt}
      \caption{The utility of the platooning system.}
      \vspace{2pt}
      \label{Fig.3}
   \end{figure}

Fig.~\ref{Fig.3} provides the utility of the whole system, including both the platooning profit kept by trucks and profit of the service provider. This result shows the effect of the pricing on the total utility of the platooning system. As we can see, the system's utility is maximized when the service provider offers the platoon coordination service to trucks for free. With the proportion of the service fee in the platooning benefit increases, the total utility of the system keeps decreasing. The system utility with $\alpha\!=\!1$ is achieved by the platoons formed spontaneously without trucks' waiting, which can also be seen in Fig.~\ref{Fig.5}.

\begin{figure}[thpb]
     \centering
     \includegraphics[width=0.98\linewidth]{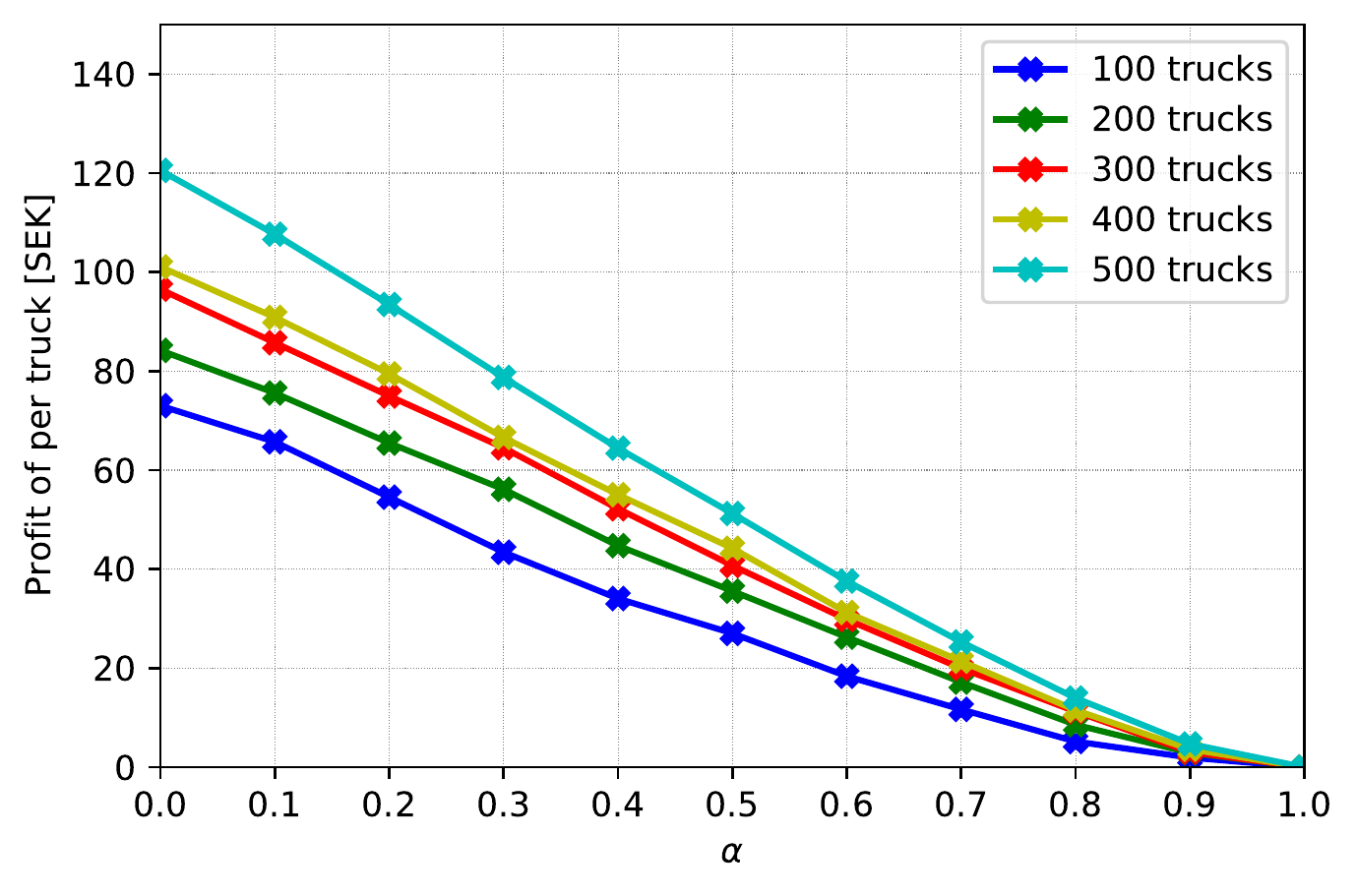}
     \vspace{-3pt}
      \caption{The average profit of per truck.}
      \label{Fig.4}
   \end{figure}

The results in Fig.~\ref{Fig.4} show the average profit of per truck in the system, i.e., the average monetary gains of every truck after paying for the service fee. It is shown that the profit of each truck trends to decrease as the parameter $\alpha$ increases, and the profit approaches to $0$ when $\alpha$ is $1$. From this figure we can also see that a platooning system with more trucks is beneficial to every individual trucks for obtaining a higher platooning profit.
\vspace{2pt}

\begin{figure}[thpb]
     \centering
     \includegraphics[width=0.94\linewidth]{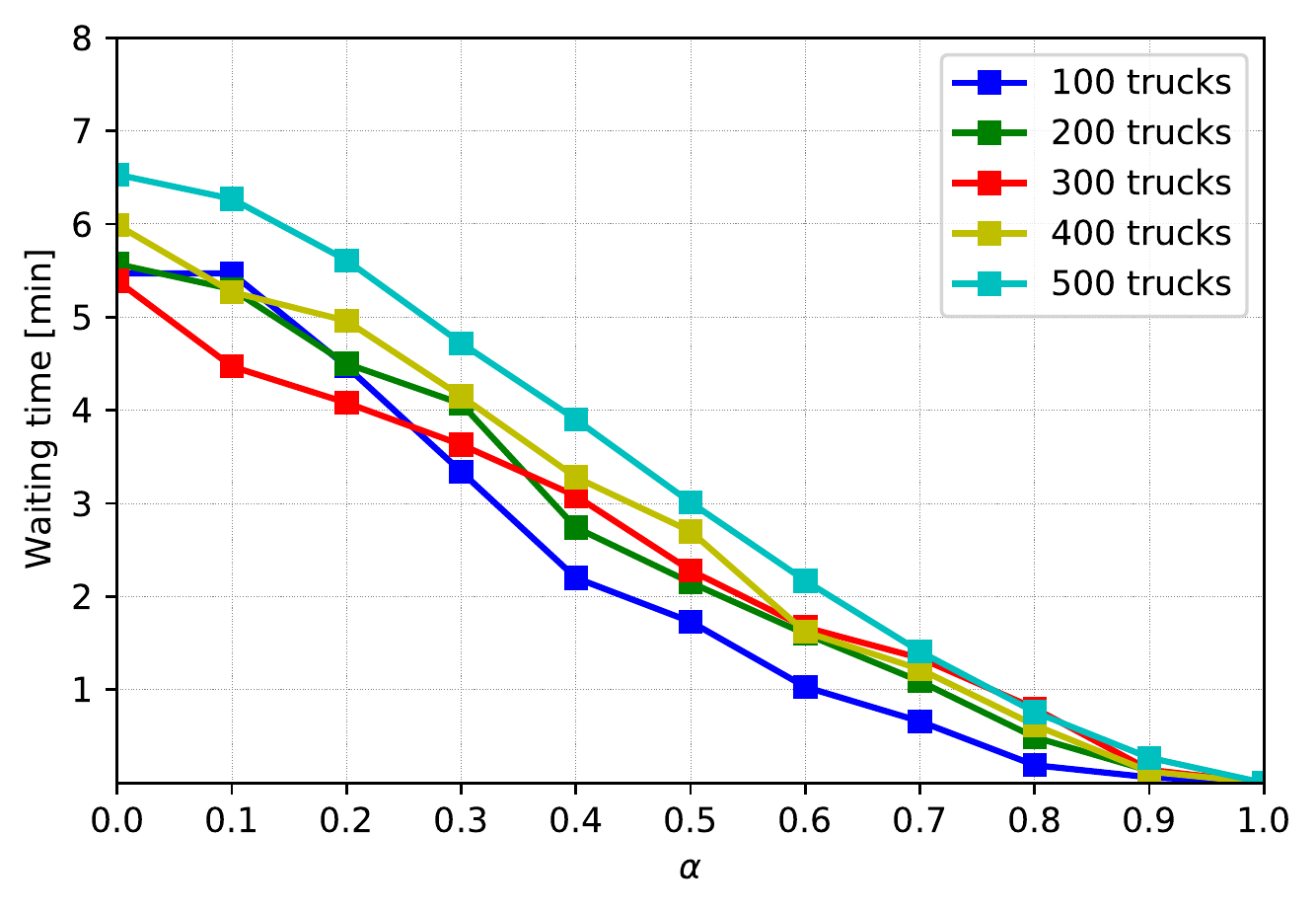}
     \vspace{-2pt}
      \caption{The average waiting time of per truck.}
      \vspace{2pt}
      \label{Fig.5}
   \end{figure}
 
The average waiting time of per truck in the whole trip is given by Fig.~\ref{Fig.5}. In the $5$ scenarios, the average waiting time of per truck is less than $7$ minutes, and no truck decides to wait when the parameter $\alpha$ equals to $1$, i.e., the service provider takes all the platooning benefit. Moreover, by comparing Fig.~\ref{Fig.3} and Fig.~\ref{Fig.5} we notice that some platoons can be formed spontaneously without coordination in particular cases. The more trucks in the system, the more profit that the system obtains from spontaneous platoons. 
\vspace{2pt}

The above simulation and evaluation results show that the third-party service provider could obtain a good platooning profit by setting a high service fee when the service provider has its monopoly in the platooning system. Otherwise, because the system's utility and the average utility of every truck are decreasing as the service fee goes up, the service provider needs to maintain a certain platooning rate by decreasing the service fee for achieving a satisfactory profit.  
\vspace{5pt}

\section{Conclusions and future directions}
This paper modelled a platooning system with a third-party service provider, which enables trucks from different carriers to use the platoon coordination service offered by the service provider without direct communication between trucks. A pricing rule was proposed for the service provider to decide the service fee for charging the follower trucks in a platoon and the compensation to charge the leader trucks. Moreover, the pricing rule was integrated with a distributed MPC based platoon coordination method, by which the service provider could compute an optimal suggested platoon solution for every truck. The pricing rule was further evaluated in the simulation over the Swedish road network with hundreds of trucks. 
\vspace{1pt}

Our simulation results show that, if the third-party service provider is in the monopoly position in the market, it can get a considerable platooning profit by setting a high service fee. If other competitive service providers exist, the service provider needs to set the pricing carefully in order to maintain trucks' willingness to use the service and achieve a satisfactory profit. In practice, a third-party service provider may have a monopoly during an early phase of the commercial platooning rollout before other actors have been established. 
\vspace{1pt}

As future work, we aim to study pricing rules when there is not a monopoly market and the trucks can choose among different service providers. Another interesting future work can be exploring new pricing rules that encourage trucks to form platoons with the financial support from governments, which helps to reduce the CO2 emissions by platooning. 
\vspace{5pt}

\section{Acknowledgment}
The work of Ting Bai, Alexander Johansson and Jonas Mårtensson was supported by the Horizon 2020 through the Project Ensemble. The work of Ting Bai was also supported by the Outstanding Ph.D. Graduate Development Scholarship from Shanghai Jiao Tong University.
\vspace{3pt}

\bibliographystyle{IEEEtran}
\bibliography{Ref}
\end{document}